\documentclass[fleqn,usenatbib]{mnras}

\usepackage{graphicx}	% Including figure files
\usepackage{newtxtext,newtxmath}
\usepackage{amsfonts}
\usepackage{appendix}
\usepackage{rotating}
\usepackage{pdflscape}
\usepackage{longtable}
\usepackage{comment}
\usepackage{xcolor}
\usepackage{orcidlink}
\usepackage[T1]{fontenc}

\DeclareRobustCommand{\VAN}[3]{#2}
\let\VANthebibliography\thebibliography
\def\thebibliography{\DeclareRobustCommand{\VAN}[3]{##3}\VANthebibliography}

\usepackage{graphicx}	% Including figure files
\usepackage{amsmath}	% Advanced maths commands
\title[Ly$\alpha$ and Physical Properties]{Ly$\alpha$ Emission Strength and Stellar Properties of Faint Galaxies from $5 < z < 8.2$}

\author[Bolan et al.]{Patricia Bolan,$^{1,\dagger}$\orcidlink{0000-0002-7365-4131},
Maru$\Breve{\textrm{s}}$a Brada$\Breve{\textrm{c}}^{1,2}$\orcidlink{0000-0001-5984-0395},
Brian C. Lemaux$^{3,1}$\thanks{Corresponding author e-mail: brian.lemaux@noirlab.edu}\orcidlink{0000-0002-1428-7036},
Victoria Strait$^{4,5}$\orcidlink{0000-0002-6338-7295},
Tommaso Treu$^{6}$\orcidlink{0000-0002-8460-0390},
\newauthor
Laura Pentericci$^{7}$,
Debora Pelliccia$^{8}$\orcidlink{0000-0002-3007-0013},
and Kelsey Glazer$^{1}$\orcidlink{0000-0002-4453-5870}, 
and Gareth C. Jones$^{9}$\orcidlink{0000-0002-0267-9024}
\\
$^{1}$Department of Physics and Astronomy, University of California, Davis, 1 Shields Ave, Davis, CA 95616, USA\\
$^{2}$University of Ljubljana, Department of Mathematics and Physics, Jadranska ulica 19, SI-1000 Ljubljana, Slovenia \\
$^{3}$Gemini Observatory, NSF’s NOIRLab, 670 N. A’ohoku Place, Hilo, Hawai’i, 96720, USA \\
$^{4}$Cosmic Dawn Center (DAWN)\\
$^{5}$Niels Bohr Institute, University of Copenhagen, Jagtvej 128, København N, DK-2200, Denmark \\
$^{6}$Department of Physics and Astronomy, University of California, Los Angeles, CA 90095-1547, USA \\
$^{7}$INAF Osservatorio Astronomico di Roma, Via Frascati 33, I-00040 Monteporzio (RM), Italy\\
$^{8}$UCO/Lick Observatory, Department of Astronomy $\&$ Astrophysics, UC Santa Cruz, 1156 High Street, Santa Cruz, CA 95064, USA \\
$^{9}$Department of Physics, University of Oxford, Denys Wilkinson Building, Keble Road, Oxford OX1 3RH, UK \\
$^{\dagger}$ Deceased
}

\date{Accepted 2024 May 21. Received 2024 May 21; in original form 2023 September 29}

\pubyear{2024}

\begin{document}
\label{firstpage}
\pagerange{\pageref{firstpage}--\pageref{lastpage}}
\maketitle

\begin{abstract}
We present a study on stellar properties of Lyman-alpha (Ly$\alpha$) emitters at 5 $< z <$ 8.2. We use 247 photometrically-selected, lensed, high-redshift, low luminosity galaxy candidates with spectroscopic follow-up. Of these, 38 are confirmed spectroscopically to be between 5 $< z <$ 8.2 via detection of Ly$\alpha$. For each galaxy and candidate, we estimate stellar mass, star formation rate, specific star formation rate, and mass-weighted age with spectral energy distribution fitting. We also measure the UV $\beta$ slope and luminosity using values from photometry. We find no strong correlation between Ly$\alpha$ equivalent width and any of these properties, as well as no significant difference between the physical properties of Ly$\alpha$ emitters and candidates without Ly$\alpha$ detected. This lack of expected trends may be explained by a combination of the evolving opacity of the IGM at these redshifts as well as the unique phase space probed by our lensed sample. Via tests on other galaxy samples which show varying strengths of correlations, we conclude that if there exist any relationships between Ly$\alpha$ EW and physical properties in the underlying population of faint galaxies, they are weak correlations. We also present the results of a spectroscopic search for CIII] emission in confirmed Ly$\alpha$ emitters at $z \sim 7$, finding no CIII] detections, but putting constraints on strong AGN activity and extreme nebular emission. 
\end{abstract}


%%%%%%%%%%%%%%%%%%%%%%%%%%%%%%%%%%%%%%%%%%%%%%%%%%

%%%%%%%%%%%%%%%%% BODY OF PAPER %%%%%%%%%%%%%%%%%%

\section{Introduction}
%Some of the biggest remaining mysteries in astronomy concern the Epoch of Reionization (EoR), which occurred roughly between the redshift range 6 $\leq z \leq$ 10. In this era, the intergalactic medium (IGM) went from being opaque, neutral hydrogen to almost entirely ionized \citep{fan_observational_2006, mcgreer_model-independent_2015, planck_collaboration_planck_2020}. 

There are still many open questions about the Epoch of Reionization (EoR) surrounding the responsible sources and the detailed mechanisms of how it occurred. Observations of high redshift ($z > 6$) galaxies are critical to answer many of these questions. Through extensive observational studies over the past decades, galaxies have been identified as likely playing a dominant role in reionization relative to other sources such as quasars \citep[e.g.][]{bouwens_reionization_2015, finkelstein_unveiling_2019, robertson_galaxy_2021, yeh_span_2023, fan_quasars_2023, robertson_identification_2023}. However, it still remains unclear whether massive and bright galaxies emitted the majority of ionizing photons coming from galaxies \citep[e.g.][]{robertson_cosmic_2015, naidu_rapid_2020, lin_empirical_2023} or the ionizing budget was dominated by fainter, yet more numerous galaxies \citep[e.g.][]{finkelstein_conditions_2019, mascia_new_2023}. 

Determining the characteristics of galaxies in the EoR through observations is crucial to constraining the relative contribution to reionization of bright and faint galaxies. With substantial samples of EoR galaxies, we can estimate physical properties and correlate those with some measure of how efficient the galaxies are at ionizing. There are a few metrics which dictate the ionizing capabilities of these early galaxies: the star formation rate, ionizing efficiency, and the rate of Lyman continuum (LyC) escape from these galaxies, $f_{\textrm{esc}}$. Some studies attempt to place constraints on the ionizing photon production rates of early galaxies, which becomes increasingly difficult with higher redshifts \citep[e.g.,][]{shivaei_mosdef_2018, nakajima_mean_2018, prieto-lyon_production_2022, saldana-lopez_vandels_2023}. Similarly, the rate of LyC escape from these galaxies is not directly attainable due to the opacity of the intergalactic medium (IGM) at these redshifts. Indirect estimations and correlations with measurable properties are possible, however. Surveys of galaxies at lower redshifts reveal a correlation between the strength of Lyman-alpha emission (Ly$\alpha$, 1216 \AA) and $f_{\textrm{esc}}$ \citep{pahl_uncontaminated_2021, flury_low-redshift_2022, begley_vandels_2022, saldana-lopez_vandels_2023, choustikov_great_2024}. Therefore, we use Ly$\alpha$ emission as a proxy for ionizing photon production in order to begin to characterize a typical galaxy that had a dominant role in reionizing the IGM. {However, we note that some studies have not found any strong correlations between Ly$\alpha$ escape and LyC escape \citep{fletcher_lyman_2019, izotov_diverse_2020, ji_hst_2020}.

Ly$\alpha$ is a powerful probe of galaxies in the EoR as it is intrinsically the strongest line in the UV. However, Ly$\alpha$ photons are scattered by neutral hydrogen, making the line a probe of the ionization state of the IGM as well as properties of the sources that emitted them \citep[e.g.][]{verhamme_3d_2006, treu_inferences_2012, dijkstra_ly_2014}. Analyzing the physical properties of Ly$\alpha$ emitters (LAEs) during the EoR, how they correlate with Ly$\alpha$ strength, and how they differ from the  general population of galaxies for which Ly$\alpha$ emission is not detected, provides a pathway to revealing the mechanisms behind the galaxies that likely powered reionization. If there is evidence of enhanced Ly$\alpha$ emission from galaxies with certain physical properties, it may be possible to distinguish which ones dominated the output of ionizing photons during reionization. 

The search for LAEs at $z > 6$ is possible due to the creation of ionized regions around these galaxies by the emission of ionizing photons from within the galaxies. Due to the resonant nature of the Ly$\alpha$ emission line, these photons are typically only detectable in galaxies with ionized regions carved out around them, favoring the detection of LAEs in overdense regions where large ionized bubbles can be created \citep{mesinger_ly_2008, endsley_mmt_2021, trapp_lyman_2023}. Because of this effect, LAEs from the EoR are excellent tracers of ionized regions in an evolving IGM.

In the past few decades, there have been increasingly more high redshift LAEs detected. With the Hubble Space Telescope ($HST$) and $Spitzer$ space telescope,  space-based photometry has unveiled many candidates high-redshift Lyman Break Galaxies (LBGs) based on multiband observations \citep[e.g.][]{stark_keck_2010, huang_spitzer_2016}. Follow-up spectroscopy of these candidates has yielded samples of confirmed LAEs. Since the successful launch and deployment of the James Webb Space Telescope ($JWST$), many more LAEs at $z > 6$ are being identified \citep[e.g.][]{jung_ceers_2023, witstok_inside_2023, scholtz_gn-z11_2023, jones_jades_2023, saxena_jades_2023, saxena_jades_2023-1, tang_jwstnirspec_2023, maseda_jwstnirspec_2023, iani_midis_2023, bunker_jades_2023}. However, while high redshift galaxies are being unveiled back to the first few hundred million years after the Big Bang, large samples of galaxies during the heart and tail-end of the EoR remain extremely useful to help characterize the drivers of reionization, especially as it is difficult to detect low equivalent width LAEs using the NIRSpec prism. Until there are substantial samples of EoR galaxies with spectroscopy from $JWST$ between $5 < z < 8$, we can gain valuable insights from the collections of galaxies with ground based spectroscopic observations available now.

In this work, we use a sample of 247 lensed $z \sim 5-8.2$ LBG candidates from \citet{fuller_spectroscopically_2020} and \citet{hoag_constraining_2019}, 38 of them with Ly$\alpha$ detected in emission. This is the largest faint ($L \sim 0.1L^*$ where $L^*$ is the characteristic luminosity) sample at this redshift, assembled from hundreds of orbits on \emph{HST}, \emph{Spitzer}, and an over six-year long campaign on the Keck DEep Imaging Multi-Object Spectrograph (DEIMOS; \citealp{faber_deimos_2003}) and Multi-Object Spectrometer For Infra-Red Exploration (MOSFIRE; \citealp{mclean_design_2010}). We estimate and compute various physical properties of the LBG candidates and spectroscopically confirmed LAEs: UV $\beta$ slope, UV luminosity, stellar mass, star formation rate (SFR), specific star formation rate (sSFR), and mass-weighted age. In order to determine if there are significant physical differences in populations of LAEs vs nonLAEs\footnote{Throughout the paper, this term will refer to LBG candidates with sufficiently constraining spectroscopic limits to rule out Ly$\alpha$ in emission at a level of EW(Ly$\alpha$)<25\AA, which is the Ly$\alpha$ strength we use to delineate LAEs from nonemitters in this work.}, we compare these properties of both samples in bulk. For the LAEs, we look at the strength of the Ly$\alpha$ emission, via both equivalent width (EW) and line luminosity, versus each estimated physical property to see if there are statistically significant trends that may help predict both the presence and strength of Ly$\alpha$ in these galaxies. Past surveys of LAEs and LBGs with this sample size typically probe luminosites of the order $L^*$ and do not extend higher than $z \sim 6$ \citep[e.g.][]{pentericci_physical_2009, stark_keck_2010, oyarzun_comprehensive_2017, santos_evolution_2020}. Due to gravitational lensing, our sample extends down to luminosites of $0.001L^*$, making this work unique in its characterization of faint EoR galaxies. 

The paper is organized as follows. In Section 2, we discuss the data and observations which comprise this sample. In Section 3, we present the methodology of the galaxy property estimation and measurement process used in this study. Section 4 describes the analysis of potential correlations between Ly$\alpha$ emission strength and physical properties of galaxies with spectroscopic confirmation of Ly$\alpha$. Section 5 covers a comparison of LAEs and the nonLAE sample. In Section 6, we present a study of spectral properties in a subsample of LAEs at $z \sim 7$. In Section 7, we present and discuss our results, and conclusions can be found in Section 8. Whenever needed, we use $\Lambda$CDM cosmology with $\Omega_{m} = 0.3, \Omega_{\Lambda} = 0.7, H_0 = 70$. All magnitudes are given in the AB system \citep{oke_secondary_1983}, and all EWs are presented in the rest frame with a positive value indicating emission.

\section{Data and Observations}
\label{sec:data}

The sample used in this study is comprised of 247 Lyman Break galaxies candidates selected using the dropout method with photometry from $HST$ and $Spitzer$. Each candidate is followed up with a spectroscopic search for Ly$\alpha$, 64 of them on Keck/MOSFIRE, and 198 on Keck/DEIMOS, with 15 of these observed with both instruments.

The data used in this analysis come from two sets of observations: a sample of galaxy candidates between $z \sim 5-7$ \citep{fuller_spectroscopically_2020}, and another set between $z \sim 7-8.2$ \citep{hoag_constraining_2019}. The candidates are detected behind massive lensing clusters. Each target LBG has photometric measurements from some combination of the following: $HST$ Advanced Camera for Surveys (ACS; \citealp{ford_advanced_1998}) and Wide Field Camera (WFC3; \citealp{kimble_wide_2008}) filters: F435W, F475W, F555W, F606W, F625W, F775W, F850LP, F814W, F105W, F110W, F125W, F140W, and F160W. Five of the clusters are from Hubble Frontier Fields (HFF, \citealp{lotz_frontier_2017}): A2744, MACS0416, MACS0717, MACS1149, and A370. Four clusters come from the Cluster Lensing and Supernova Survey with Hubble (CLASH, \citealp{postman_cluster_2012}), MACS0744, MACS1423, MACS2129, and RXJ1347, and the last, MACS2214, has \emph{HST} imaging from the Spitzer UltRa Faint SUrvey Program (SURFSUP, \citealp{bradac_spitzer_2014}). In addition to \emph{HST} imaging from these programs, each cluster has \emph{Spitzer} observations from SURFSUP and the \emph{Spitzer} HFF programs from the 3.6$\mu$m and 4.5$\mu$m channels on the Infrared Array Camera (IRAC; \citealp{fazio_infrared_2004}). A summary of the cluster fields in this sample is given in Table \ref{table:obs_1}. 

The original photometric sample from which the smaller spectroscopic sample was chosen was selected by the Lyman Break technique, as mentioned above. To choose the candidates included in this work, we use constraints on the photometric redshift and probability of redshift distribution, or $P(z)$. These $P(z)$s are determined using Easy and Accurate Redshifts from Yale (EA$z$Y; \citealp{brammer_eazy_2008}). This process is described in detail in \citet{huang_detection_2016} and \citet{strait_stellar_2020}. Briefly, the code performs $\chi^2$ minimization over a grid of redshifts and computes the $P(z)$ distribution assuming a flat prior due to the candidates being lensed. The sample of candidate LAEs to be followed up spectroscopically was selected following the methods of \citet{hoag_constraining_2019} and \citet{fuller_spectroscopically_2020}.

\begin{table*}Summary of Observations
    \begin{center}
    \begin{tabular}{||c c c c c c c c||} 
     \hline
    Cluster Name & Short Name & $\alpha_{J2000}$ (deg) & $\delta_{J2000}$ (deg) & $z_{\textrm{cluster}}$ & $\#$ of candidates & \multicolumn{1}{p{2cm}}{\centering $\#$ of Ly$\alpha$ \\ detections} & \multicolumn{1}{p{3cm}}{\centering \emph{HST} Imaging \\ (\emph{Spitzer} Imaging)} \\ 
    \hline\hline
    Abell 2744 &  A2744 & 3.5975000 & $-$30.39056 & 0.308 & 25 & 1 & HFF\\ 
    Abell 370 & A370 & 39.968000 & $-$1.576666 & 0.375 & 28 & 3 & HFF\\
    MACSJ0416.1$-$2403 & MACS0416 & 64.039167 & $-$24.06778 & 0.420 & 21 & 3 & HFF/CLASH\\
    %MACSJ0454.1-0300 & MACS0454 & 73.545417 & -3.018611 & 0.540 &  HST-GO-11591/GO-9836/GO-9722 (SURFSUP)\\
    MACSJ0717.5+0745 & MACS0717 & 109.38167 & 37.755000 & 0.548 & 10 & 5 & HFF/CLASH (SURFSUP) \\
    MACSJ0744.8+3927 & MACS0744 & 116.215833 & 39.459167 & 0.686 & 24 & 3 & CLASH (SURFSUP)\\
    MACSJ1149.5+2223 & MACS1149 & 177.392917 & 22.395000 & 0.544 & 27 & 2 & HFF/CLASH (SURFSUP)\\
    MACSJ1423.8+2404 & MACS1423 & 215.951250 & 24.079722 & 0.545 & 28 & 6 & CLASH (SURFSUP)\\
    MACSJ2129.4$-$0741 & MACS2129 & 322.359208 & $-$7.690611 & 0.568 & 30 & 5 & CLASH (SURFSUP)\\
    MACSJ2214.9$-$1359 & MACS2214 & 333.739208 & $-$14.00300 & 0.500 & 11 & 1 & SURFSUP\\
    %RCS2-2327.4-0204 & RCS2327 & 351.867500 & -2.073611 & 0.699 & HST-GO-10846 (SURFSUP) \\
    RXJ1347.5$-$1145 & RXJ1347 & 206.87750 & $-$11.75278 & 0.451 & 43 & 9 & CLASH (SURFSUP)\\ [1ex] 
     \hline
    \end{tabular}
    \caption{Summary of observations for 10 lensing clusters used in this work. }
    \label{table:obs_1}
    \end{center}
\end{table*}

\subsection{Photometry}
\label{sec:phot}

The photometric measurements for the candidates in the cluster fields of A2744, A370, MACS0416, MACS0717, and MACS1149 are from the ASTRODEEP team \citep{castellano_astrodeep_2016, di_criscienzo_astrodeep_2017, bradac_hubble_2019}. For the remaining clusters, we use an identical method to that employed by the ASTRODEEP team for photometric measurements. Briefly, point-spread function (PSF) matched HST images were created, in which all of the HST images had their PSF degraded to match that of the F160W images. The F160W image was used as the detection band for all fields. In order to improve the detection of faint objects, intracluster light (ICL) was subtracted for targets in each cluster field except for MACS0744 and MACS2214. In these two fields, the ICl subtraction was not performed because the high-redshift objects in these clusters were not heavily contaminated by the ICL. HST photometry was then measured using Source Extractor (SExtractor; \citealp{bertin_sextractor_1996}) in dual-image mode with F160W as the detection image. Photometry for $Spitzer$ images was extracted using \textsc{T-PHOT} \citep{merlin_t-phot_2015}. Further details on this process can be found in \citet{huang_spitzer_2016} and \citet{fuller_spectroscopically_2020}. 

%All candidates in the sample are selected based on the Lyman break technique. 

\subsection{Spectroscopy}
\label{sec:spec} 

Out of 247 LBGs, 38 are confirmed LAEs, presented in \citet{fuller_spectroscopically_2020, hoag_constraining_2019}. To select the final sample used in this work, an inclusive cut of at least 1\% of the total integrated probability density of the redshift, $P(z)$, lying in the redshift range where Ly$\alpha$ could be detected on the given instrument with the given setup is used. All analyses done using LBG candidates in this paper are weighted according to amount of the $P(z)$ distribution which lies in this range for a given instrument. This process is described further in section~\ref{sec:galprops}. The spectroscopic observations were made between 2013 and 2017. The average 3$\sigma$ observed flux limit for the nonemitters is$~\sim 2 \times 10^{-18} \ \textrm{erg/s/cm}^{2}$, providing deep constraints on Ly$\alpha$ EW for those targets which did not have detected emission lines.  More details on the observations specifics and conditions can be found in \citet{hoag_constraining_2019} and \citet{fuller_spectroscopically_2020}.

As the targets in this survey are gravitationally lensed, determination of magnification, or $\mu$, values were determined for each candidate which we observe. The lens models used are described by \citet{bradac_strong_2005, bradac_focusing_2009}, with details on models for each individual cluster described in Section 3.2 of \citet{hoag_constraining_2019}. Magnification values for each LBG are determined using the photometric redshift, position, and the generated best-fit magnification map \citep[e.g.][]{hoag_grism_2016, finney_mass_2018}. Over the entire sample, magnification values span three orders of magnitude, from $\sim$ 1 to $\sim$ 200. 

\section{Methods}
\label{sec:methods}
There are six physical properties beyond Ly$\alpha$ emission which we focus on for this study; two are determined directly from the photometry without the use of models other than those used to determine the photometric redshift ($z_{\textrm{phot}}$): UV $\beta$-slope and UV luminosity. The remaining four properties are estimated using spectral energy density (SED) fitting: stellar mass, star formation rate, specific star formation rate, and mass-weighted age. For each of these properties, we look at differences between LAEs and nonLAEs, as well as how these properties correlate with Ly$\alpha$ strength for those targets with spectroscopic detections. 

\subsection{Ly$\alpha$ EW}
For the 38 galaxies with Ly$\alpha$ detections in the sample, we use the EW values, or relative strength of the Ly$\alpha$ emission line compared to the continuum of the galaxy. What constitutes a detection, as well as the calculation of Ly$\alpha$ EW values, are described by \citet{hoag_constraining_2019} and \citet{fuller_spectroscopically_2020}. While we do not detect emission in the majority of our sample, we are still able to constrain the upper 1$\sigma$ limit on Ly$\alpha$ EW via 

\begin{equation}
\label{eq:EW}
\textrm{EW}_{lim} = \frac{f_{\textrm{lim}}(\lambda)}{f_{\textrm{cont}}(1+z)}
\end{equation}
where the continuum flux density, $f_{\textrm{cont}}$, is defined as

\begin{equation}
\label{eq:3}
f_{\textrm{cont}} = 10^{-0.4(m_\textsc{AB}+48.6)}c/\lambda^2 \ 
$\textrm{erg/s/cm}$^2/$\AA$
\end{equation}
computed using continuum flux redward of the expected Ly$\alpha$ line, using the following HST bands for $m_\textsc{AB}$. For galaxies observed by DEIMOS at $5 < z < 7$, we typically use the apparent magnitude in the F105W band, which corresponds to an average rest-frame wavelength of $\lambda \sim$ 1500 \AA. If there is no F105W data available, F125W (rest-frame $\lambda \sim$ 1800 \AA) is used, and, in a few cases, where there is neither an F105W nor F125W magnitude, F140W (rest-frame $\lambda \sim$ 2000 \AA) is used. For those observed with MOSFIRE between $7 < z < 8.2$, we use the F160W band. The quantity $f_{\textrm{lim}}$ is the flux density value of the 1$\sigma$ noise spectrum at the expected spectral location of the emission line. We note that gravitational lensing is achromatic, and therefore EW is invariant with respect to magnification value. 

\subsection{UV Property Calculations}
\label{sec:UVcalcs}
Two UV properties of the galaxies in our sample can be calculated via photometric fluxes: the UV $\beta$ slope and UV luminosity. The UV continuum slope of a galaxy's spectrum, or its $\beta$ slope, characterizes its flux redward of Ly$\alpha$ emission with the relation $f_{\lambda} \propto \lambda^{\beta}$. The steepness of the $\beta$ slope can give insight into stellar populations and the degree of dust reddening in a galaxy \citep{buat_goods-_2012, yamanaka_uv_2019, calabro_vandels_2021, chisholm_far-ultraviolet_2022}. We compute the $\beta$ slope for each LBG candidate that has the requisite data as set by the following. We require at least two magnitude measurements in filters redward of the expected Ly$\alpha$ emission. All $\beta$ slopes are calculated in the observed frame using linear regression fitting of the uncertainty-weighted magnitude values against the effective wavelength of the filter from \textit{Scikit-Learn} \citep{pedregosa_scikit-learn_2018}. 

The intrinsic UV luminosity, or $M_\textsc{uv}$, is calculated via
\begin{multline}
       M_\textsc{uv} \approx M_\textsc{fuv} = m_{F160W} + 2.5\log_{10}(\mu) - \\
         5(\log_{10}(d_{L})-1) + 2.5\log_{10}(1+z) +0.12
\end{multline}
where $m_{F160W}$ is the apparent magnitude in the F160W band, $\mu$ is the median magnification value recovered from lens modelling, $d_L$ is luminosity distance in parsecs, $z$ is the redshift, and 0.12 is a K-correction to correct to rest frame 1600\AA \ (see \citealp{fuller_spectroscopically_2020} for details). 

Figure~\ref{fig:z_muv} shows the spectroscopically confirmed redshifts as a function of $M_\textsc{uv}$, color coded by Ly$\alpha$ EW for each of the confirmed LAEs. For more details on the full sample, including plots of the reduced spectra and Ly$\alpha$ detections, signal to noise values, and quality flags, we refer the reader to \citet{fuller_spectroscopically_2020} and \citet{hoag_constraining_2019}.

\begin{figure}
\includegraphics[width=9.2cm]{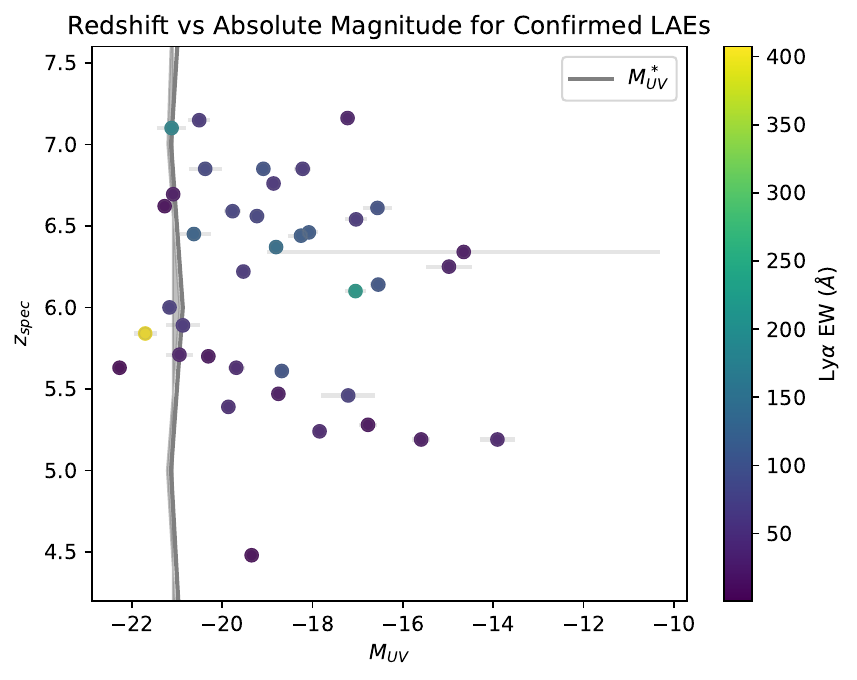}
\caption{Spectral $z$ values as a function of absolute UV magnitude for all confirmed LAEs in the sample. The scale bar indicates the Ly$\alpha$ EW value. The gray vertical line indicates the characteristic UV magnitude value per redshift bin from \citet{bouwens_z_2022}, showing that our sample is mostly fainter than the characteristic brightness.}
\label{fig:z_muv}%
\end{figure}

\subsection{Estimating Galaxy Properties}
\label{sec:galprops}
To estimate stellar properties of the sample, we use Bayesian Analysis of Galaxies for Physical Inference and Parameter EStimation (BAGPIPES, \citealp{carnall_inferring_2018}). BAGPIPES fits physical parameters using the MultiNest sampling algorithm \citep{feroz_multimodal_2008, feroz_multinest_2009}. We use the default set of stellar population templates from Bruzal and Charlot \cite[BC03]{bruzual_stellar_2003}. The SED fitting is done using the initial mass function (IMF) from \citet{kroupa_mass_2002}, a metallicity of 0.02$Z_{\sun}$, a Calzetti dust law \citep{calzetti_dust_2000}, and a constant star formation history (SFH). We allow dust extinction to range from $A_v$ = 0-3 magnitudes. When we allow metallicity to vary between 0-2$Z_{\sun}$, our median best fit value is 0.02$Z_{\sun}$, which is broadly consistent with metallicities estimated from other sets of observations and simulations for samples of galaxies at similar redshifts and stellar masses to our own (e.g., \citealt{dekel_efficient_2023, hirschmann_metallicity_2023, nakajima_jwstmetals_2023, seeyave_first_2023, curti_jwstmetals_2024}). 
We find no significant difference in the resultant posteriors whether we fix metallicity at 0.02$Z_{\sun}$ (to reduce computational time) or allow it to vary. The choice of 0.02$Z_{\sun}$ was selected from a sample of fiducial redshifts as the closest to the median when allowing metallicity to be a free parameter. We also perform parallel runs employing a delayed tau SFH and find that it has no effect on our final conclusions. While the distribution of stellar properties change with the selected SFH, the aggregate comparison and correlation study results remain unchanged. We also run the fits using different dust laws - Cardelli, CF00 \citep{charlot_simple_2000}, and Salim \citep{salim_dust_2018} - and find no significant difference in the resultant estimated parameters. For each SED run, we input a redshift value to fit at. For targets with Ly$\alpha$ detections, the redshift is fixed either at the spectroscopic redshift or, in the case of nonLAEs, the redshift sampled from the $P(z)$ for a given Monte Carlo (MC) iteration through a process we describe in the next section.

\subsection{Monte Carlo Processes and Weighting Scheme}
\label{sec:mc}

\begin{figure}
\includegraphics[width=9cm]{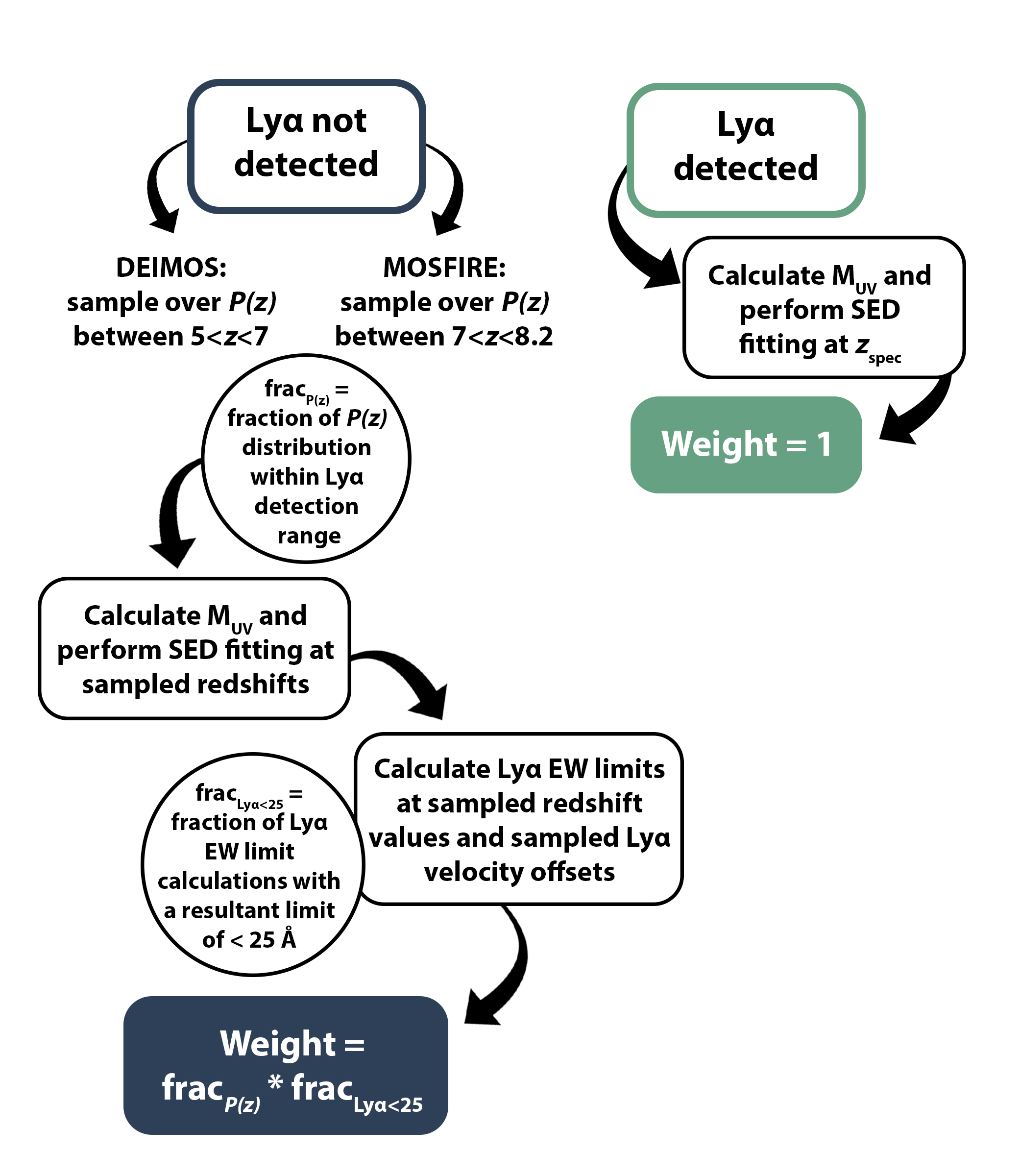}
\caption{Schematic diagram explaining the process used to determine weights used in the analysis, incorporating both the fraction of the $P(z)$ distribution which lies within the Ly$\alpha$ detection range and the fraction of EW limit calculations which are $\leq$ 25\AA \ at the $3\sigma$ level. This method allows us to take into account the uncertainties in redshift for nonLAEs.}
\label{fig:MC}%
\end{figure}

The determination of all of these physical properties relies on the input of a redshift value for the galaxy in question; parameters estimated from SED fitting, $M_\textsc{uv}$, $\beta$ slope, and Ly$\alpha$ EW limit all have dependence on $z$. To determine these values for LBGs which do not have a confirmed spectroscopic redshift, we use an MC sampling method in order to properly treat the uncertainty in redshift. Figure~\ref{fig:MC} provides a schematic diagram of the MC processes and how they are used to determine appropriate weights for properties of nonemitters when comparing their properties to LAEs. We give detailed descriptions of these processes below. 

In comparing properties of LAEs against those of nonLAEs, we create a weighting system which takes into account the uncertainty in redshift and how that propagates to uncertainties in SED-derived properties and Ly$\alpha$ EW measurements. For physical properties, we sample over the part of the $P(z)$ distribution which is in the range of possible Ly$\alpha$ detection. For DEIMOS targets, the Ly$\alpha$ line is potentially visible for targets between $5 < z < 7$, and for those with spectroscopy from MOSFIRE, between $7 < z < 8.2$. After sampling a distribution of $z$ values from each target's $P(z)$, we run the SED fit at the sampled redshifts and produce distributions of physical properties for each of the nonemitters. When we compare these properties for LAEs against nonemitters, we include the full distribution of output parameters for each galaxy, with each value weighted as described below. 

As the flux limits from our detections are wavelength dependent, the redshift also affects the determination of Ly$\alpha$ EW from the expected spectral location of the emission line. We once again sample redshifts from the $P(z)$ in the desired redshift range. We then sample from the distribution of Ly$\alpha$-interstellar medium (ISM) velocity offsets ($\Delta v$) from \citet{cassata_alpine-alma_2020}, as Ly$\alpha$ emission is often spectrally offset from the systemic redshift of a galaxy. For a chosen redshift, we determine the spectral location of Ly$\alpha$, apply the offset sampled from the Ly$\alpha$ $\Delta v$ distribution, and choose the value in the flux density spectrum at that wavelength. We then calculate the 1$\sigma$ Ly$\alpha$ EW upper limit using Eq.~\ref{eq:EW} and multiply it by 3 to obtain 3$\sigma$ upper limits. 

When comparing properties of LAEs and nonLAEs, we use the fiducial cutoff of  EW$_{\textrm{Ly}\alpha}<$ 25\AA \ to determine non-LAEs \citep[e.g.][]{mason_universe_2018, pentericci_candelsz7_2018}. Once we obtain a distribution of output parameters from the SED fitting as well as one of measured Ly$\alpha$ EW limits for each nonemitter, we create a weight for each one based on how likely they are to be in the the desired redshift range and have an EW limit < 25\AA. The nonemitter sample is weighted by the product of the fraction of EW determinations runs for which Ly$\alpha$ EW is $< 25$ \AA \ and the fraction of total integrated $P(z)$ which is in the desired range for the given spectroscopic instrument. As each nonemitter has a distribution of 100 values for the properties estimated from SED fitting, each of those values receives a weight of 1/100th of that of the total weight for the galaxy. 
When comparing properties of LAEs and nonLAEs, we use the fiducial cutoff of  EW$_{\textrm{Ly}\alpha}<$ 25\AA \ to determine non-LAEs \citep[e.g.][]{mason_universe_2018, pentericci_candelsz7_2018}. Once we obtain a distribution of output parameters from the SED fitting as well as one of measured Ly$\alpha$ EW limits for each nonemitter, we create a weight for each one based on how likely they are to be in the the desired redshift range and have an EW limit < 25\AA. The nonemitter sample is weighted by the product of the fraction of EW determinations runs for which Ly$\alpha$ EW is $< 25$ \AA \ and the fraction of total integrated $P(z)$ which is in the desired range for the given spectroscopic instrument. As each nonemitter has a distribution of 100 values for the properties estimated from SED fitting, each of those values receives a weight of 1/100th of that of the total weight for the galaxy. As an example, if a candidate observed with DEIMOS has 40$\%$ of its $P(z)$ distribution within $5<z<7$ and 80$\%$ of the Ly$\alpha$ EW limit calculations below a 3$\sigma$ limit of 25\AA\, then each individual realization of that galaxy's properties would have a weight of $0.4*0.8$/100 = 0.0032, and that galaxy's total contribution to the distribution would have a weight of 0.32.

\section{Ly$\alpha$ EW vs Physical Properties}
\label{sec:EWvsphys}
For the 38 galaxies which have spectroscopic Ly$\alpha$ detections, we use the EW values calculated by \citet{hoag_constraining_2019} and \citet{fuller_spectroscopically_2020} and see if there are trends between a galaxy's Ly$\alpha$ EW and the following physical properties: stellar mass, SFR, sSFR, mass-weighted age, UV $\beta$ slope, and $M_\textsc{uv}$. We also perform the same analysis with delensed Ly$\alpha$ line luminosity rather than EW and recover the same results for all properties except $M_\textsc{uv}$, an exception we discuss in Section~\ref{sec:uvprops}. 

There have been many studies looking at these correlations at $2 \leq z \leq 7$ \citep[e.g.,][]{pentericci_physical_2009, kornei_relationship_2010, nilsson_nature_2011, hathi_vimos_2016, oyarzun_comprehensive_2017, du_redshift_2018, marchi_vandels_2019, santos_evolution_2020, pucha_lyman-alpha_2022, mccarron_stellar_2022, reddy_effects_2022, napolitano_identifying_2023, ortiz_introducing_2023, saldana-lopez_vandels_2023, jones_jades_2023}. The outcomes of these studies are varied and show a range of different, sometimes conflicting, trends, depending on the property being studied. In the following sections, we present the results of our work and discuss how they compare to others. We perform Spearman rank tests to quantify any correlations between each of these properties and the Ly$\alpha$ EW. One outcome parameter of the test is a $\rho$ value which ranges from -1 to 1 and quantifies the strength of the correlation, with -1 and 1 designating monotonic anticorrelation and correlation, respectively. The p-value gives a measure of the significance of the correlation, with a value less than 0.005 indicating a correlation at $\geq 3\sigma$ significance. Figure~\ref{fig:LAEprops} shows the Ly$\alpha$ EW vs each of these physical properties along with the results from the corresponding correlation test. Below we present the results of our study; discussions on how they vary from other surveys and potential reasons why can be found in Section~\ref{sec:discussion}.

\begin{figure*}
\includegraphics[width=18cm]{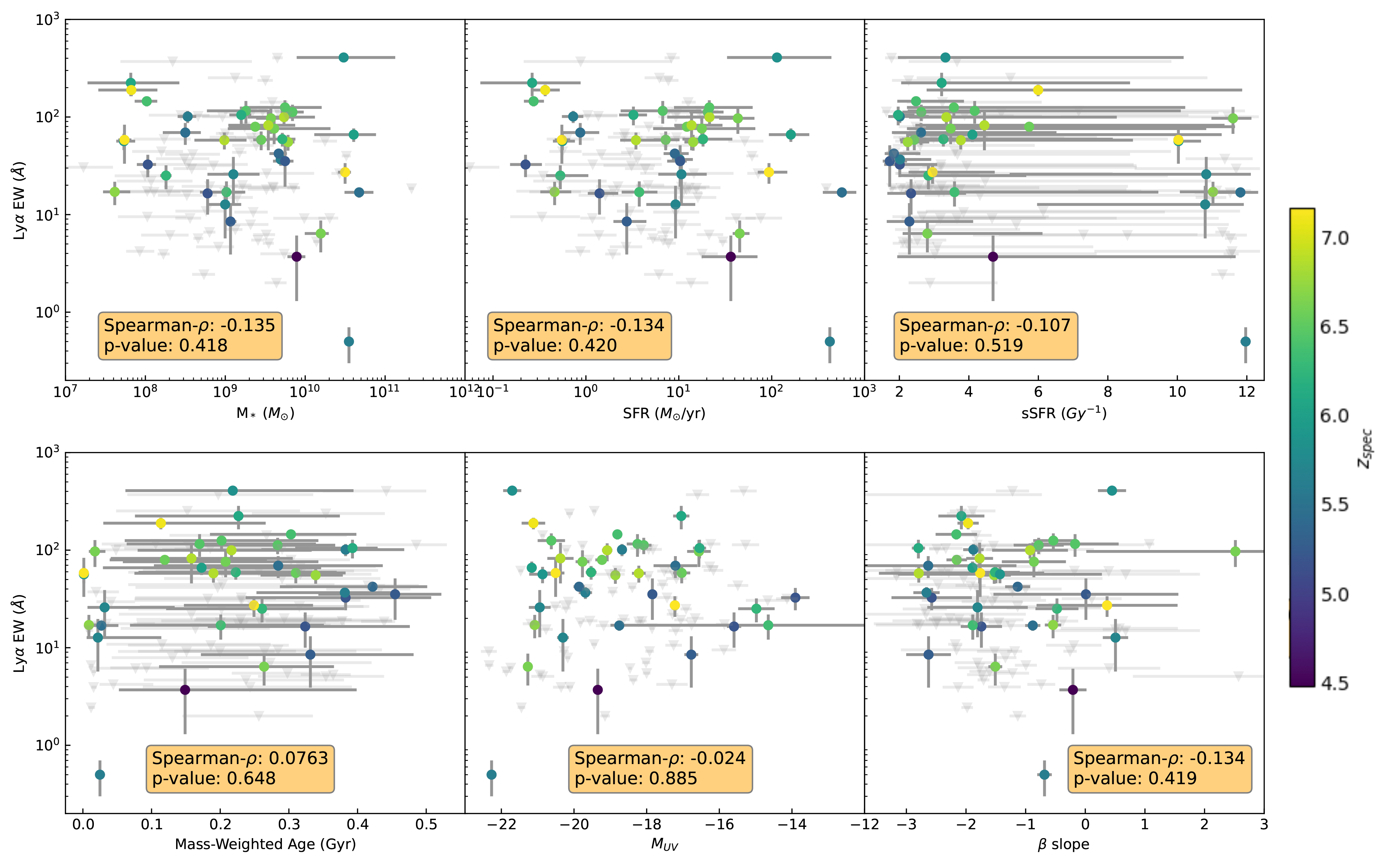}
\caption{Ly$\alpha$ EW vs various stellar and UV properties: stellar mass ($M^*$), SFR, sSFR, mass-weighted age, $M_\textsc{uv}$, and UV $\beta$ slope. The Spearman correlation coefficient ($\rho$) and p-value are shown as insets for each corresponding property. We show upper limits from nonemitters which have at least 50$\%$ of their $P(z)$ within the Ly$\alpha$ detection range as faint, inverted grey triangles. We note that while these are plotted here, upper limit values do not enter into our correlation tests. While there is some evidence of anticorrelation between Ly$\alpha$ EW and physical parameters, none of them are statistically significant.}
\label{fig:LAEprops}%
\end{figure*}

\subsection{UV Properties}
\label{sec:uvprops}

We first look at Ly$\alpha$ EW vs $\beta$ slope (see bottom right panel of Figure~\ref{fig:LAEprops}), finding no significant relationship. From previous studies and galactic physics, we may expect that larger Ly$\alpha$ EW would correlate with more negative (bluer) $\beta$ slopes. As $\beta$ provides insight into dust attenuation (although with some nuances $-$ see \citealp{alvarez-marquez_dust_2016} and references therein), galaxies with more negative slopes may be expected to have higher EW values, as dust extinction plays a key role in hindering Ly$\alpha$ escape from galaxies \citep{blanc_hetdex_2011, hagen_spectral_2014}. However, we do not find any statistically significant relationship between $\beta$ slope and Ly$\alpha$ EW ($\rho$ = -0.134, p-value = 0.419).

Next we explore how UV luminosity may relate to Ly$\alpha$ EW values. Other studies at $2<z<7$ typically find that fainter galaxies tend to have larger EW(Ly$\alpha$) \citep{stark_keck_2010, jones_keck_2012, oyarzun_comprehensive_2017, du_redshift_2018, santos_evolution_2020, jones_jades_2023}. The most similar sample is the recent one from $JWST$ Advanced Deep Extragalactic Survey (JADES)  with spectroscopy from $JWST$ NIRspec presented by \citet{jones_jades_2023}, who find that the apparent correlation between Ly$\alpha$ EW and $M_\textsc{uv}$ found in their sample is not necessarily intrinsic, but may be due to a flux sensitivity limit in the survey. It does not have low-luminosity galaxies with EW values in the 10-100\AA \ range, where our sample does include some of these. The apparent anticorrelation may be expected, as brighter galaxies typically have evolved to larger stellar masses, which is correlated with higher dust content, leading to the destruction of Ly$\alpha$ photons, and an older stellar population \citep[e.g.][]{silva_modeling_1998}. Similarly, other studies have found an increase in the escape fraction of Ly$\alpha$ in galaxies with lower UV luminosities \citep{prieto-lyon_production_2022, saldana-lopez_vandels_2023, mascia_closing_2023}. These may suggest that UV fainter galaxies have stronger Ly$\alpha$ emission. However, our results show no significant relationship between the two properties for our sample ($\rho$ = -0.024, p-value = 0.885 $-$ see Figure~\ref{fig:LAEprops}). We also compare the delensed Ly$\alpha$ line luminosity to the intrinsic UV luminosity and find strong and significant anticorrelation ($\rho = -0.591$, p-value=$9.22e^{-5}$)., i.e. more intrinsically UV luminous galaxies exhibit Ly$\alpha$ emission that is significantly stronger than their UV fainter counterparts. Such a trend is expected in our data given the lack of a significant relationship between EW(Ly$\alpha$)-$M_\textsc{uv}$ among LAEs presented in this study. We note that the large upper limits on Ly$\alpha$ EW for some of the nonLAEs can be attributed to a combination of faint UV magnitudes, relatively low exposure times, and poor observing conditions.

\subsection{Stellar Properties}
\label{sec:mass}

The properties considered in this section concern stellar populations and are determined from SED fitting. We find no evidence of any significant relationships between these SED-fit properties and EW(Ly$\alpha$) in our sample, as can be seen in Figure~\ref{fig:LAEprops}. Along the same line of reasoning as why UV fainter galaxies may have stronger Ly$\alpha$ EW, previous studies have found stronger EW values in relatively lower mass galaxies between $2<z<6$. \citet{pentericci_physical_2007, blanc_hetdex_2011, nilsson_nature_2011, hagen_spectral_2014, oyarzun_how_2016, oyarzun_comprehensive_2017, du_redshift_2018, pucha_lyman-alpha_2022} all report an anticorrelation between EW and stellar mass to varying degrees of significance. However, \citet{kornei_relationship_2010} and \citet{hathi_vimos_2016} do not find any significant correlation between the two quantities in their samples, although \citet{kornei_relationship_2010} observations are missing near-IR photometry to accurately determine masses.

%\citet{du_redshift_2018} find that the anticorrelation becomes stronger with increased redshift from $z \sim 2$ to $z \sim 4$.

The relationships found in literature between star formation rates and Ly$\alpha$ EW generally points toward less star-forming galaxies having larger EW values. \citep[e.g.][]{kornei_relationship_2010, hathi_vimos_2016, oyarzun_comprehensive_2017, trainor_predicting_2019, ortiz_introducing_2023}. However, \citet{marchi_vandels_2019} and \citet{pucha_lyman-alpha_2022} do not find any correlation between EW and SFR in galaxies between $2.5<z<4.5$. Our results align with the latter cases, as we find no significant evidence that the two parameters are related. We also estimate specific star formation rates and how Ly$\alpha$ EW depends on it, and do not find any statistically significant relationship between sSFR and Ly$\alpha$ emission strength (see Figure~\ref{fig:LAEprops} upper middle and right panels). 

%It has been found that galaxies from the first billion years of the Universe's existence tend to having higher specific star formation rates (sSFRs) than those in the local universe \citep{dekel_formation_2009, fakhouri_merger_2010, sparre_star_2015, topping_alma_2022}.

The relationship between age and Ly$\alpha$ EW has perhaps the most varied results in the literature. We find no relationship between mass-weighted age and Ly$\alpha$ EW in our sample, but also note the large errors on some age values, which can be seen in Figure~\ref{fig:LAEprops}. Some studies spanning $2 < z < 6$ have found that higher EWs are found in older galaxies \citep{kornei_relationship_2010, marchi_vandels_2019, mccarron_stellar_2022}, some have found the opposite trend \citep{pentericci_physical_2007, santos_evolution_2020, reddy_effects_2022}, and others find no correlation \citep{pentericci_physical_2009}. Ages are also highly dependent on the SFH chosen during SED fitting. As mentioned earlier, we also estimate the ages for this sample using a delayed-$\tau$ SFH. While the ages are consistently lower for a delayed-$\tau$ SFH relative to those derived using a constant SFH, there still remains no significant relationship between Ly$\alpha$ EW and mass-weighted age. The previously discussed properties are also affected by the choice of SFH in the SED fitting. Overall, the stellar mass and SFR values increase when we use a delayed SFH rather than constant with the median SFR value over twice as high with the delayed SFH compared to constant. However, as discussed in ~\ref{sec:galprops}, we still find no significant correlations between Ly$\alpha$ EW and any of the properties. We discuss possible physical and statistical reasons behind the lack of significant correlations in our sample in depth in Section~\ref{sec:discussion}. For all the properties, in addition to the p-values showing insignificant relationships, the absolute value of $\rho$ is small in all cases, such that even if there was a significant correlation or anticorrelation, it would be weak.

\section{Comparing LAEs and Nonemitters}
\label{sec:LAEvsnon}

\begin{figure}
\includegraphics[width=8.6cm]{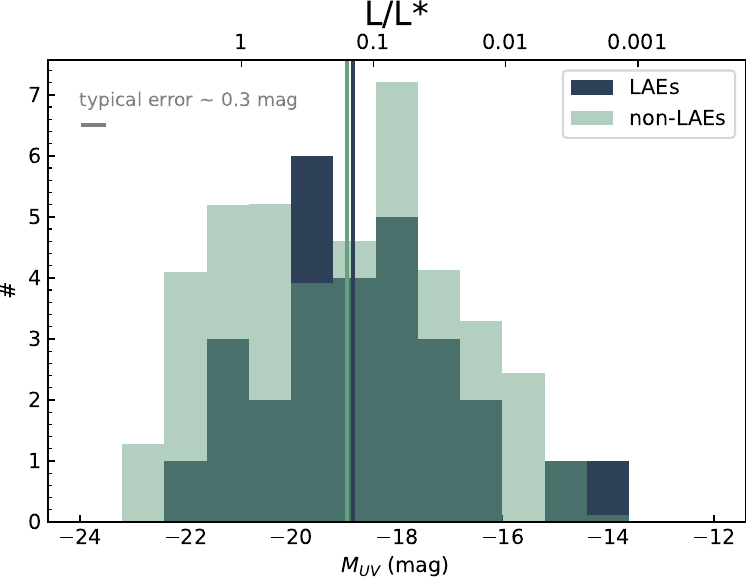}
\caption{Distribution of intrinsic UV luminosity for LAEs vs nonemitters, calculated on the peak of the photometric redshift distribution for nonemitters if it falls within the Ly$\alpha$ detection range, and the center of the range if not. Values for the nonemitter population are weighted by the product of the fraction of $P(z)$ within the range where Ly$\alpha$ could be detected and the fraction of EW calculation MC iterations for which the $3\sigma$ EW limit is $\leq 25$\AA.}
\label{fig:muv}%
\end{figure}

\begin{figure}
\includegraphics[width=8.6cm]{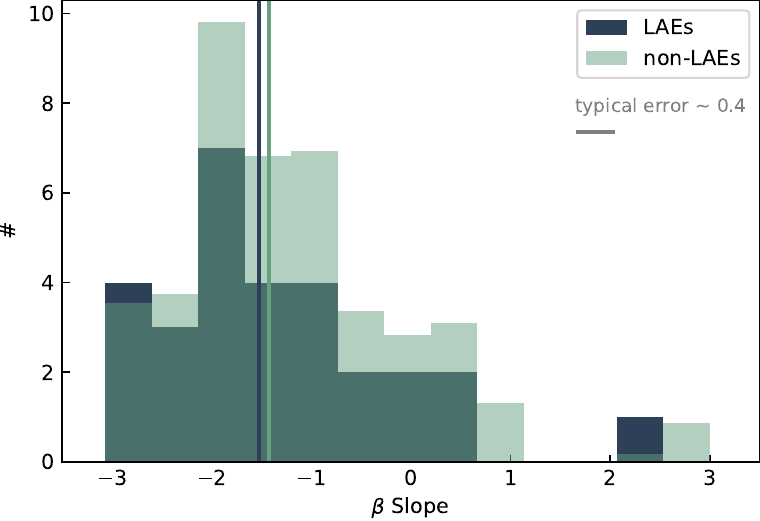}
\caption{Distribution of UV $\beta$ slope for LAEs vs nonemitters. We note that the LAE with a slope > 2 only had photometric data from two filters (F140W and F160W) and both detections were at the edge of the observational limit ($m_\textsc{ab} \sim$ 29 with large errors). Positive values for nonLAEs either have a similar situation with only two photometric fluxes or could be the effect of low-$z$ interlopers which are downweighted but still enter the distribution. Values for nonLAEs are weighted using the process described in Section~\ref{sec:mc}.}
\label{fig:beta}%
\end{figure}

In order to gain insight into what may be driving the escape of Ly$\alpha$ photons in LAEs, we also compare the distributions of the six physical properties explored above for both LAEs and nonLAEs, using the cutoff of Ly$\alpha$ EW $\geq$25\AA \ to define the LAE sample. Galaxies which have detected Ly$\alpha$ emission measured at < 25\AA \ EW are also placed into the nonemitter sample. While there are slight deviations in the distributions, we do not find any significant differences between any of the six properties of LAEs and nonemitters based on the results of both two-sample Kolmogorov-Smirnov (KS) tests and Mann-Whitney U tests. Previous studies have found statistically significant differences between the populations of LAEs and nonemitters \citep[e.g.][]{pentericci_physical_2007, stark_keck_2010, jones_keck_2012, napolitano_identifying_2023}. However, others such as \citet{hathi_vimos_2016} generally do not.  Very broadly, LAEs have been found in past works to be generally lower mass, fainter, and have less dust and bluer $\beta$ slopes than nonLAEs, but there are some inconsistent results across different surveys. Potential explanations for our lack of significant differences are discussed further in Section~\ref{sec:discussion}.

We compare the $M_\textsc{uv}$ distributions of emitters and nonemitters. As discussed earlier, candidates without spectroscopic confirmation have some uncertainty in their $M_\textsc{uv}$, dominated by the uncertainty in $z$ and consequently $d_L$. To mitigate this when comparing the luminosities of LAEs and nonemitters, we perform a similar MC sampling as described in Section~\ref{sec:mc}. In KS tests comparing the $M_\textsc{uv}$ distributions of LAEs and nonemitters in this iterative process, a typical p-value is between 0.2 and 0.4, with none below 0.005, and thus, we do not have sufficient evidence to reject the null hypothesis that the two samples are drawn from the same .

In Figure~\ref{fig:muv}, we show the distribution of $M_\textsc{uv}$ values for the LAEs against those of the nonemitters, calculated from the peak $z_{\textrm{phot}}$ value if it is within the desired redshift range, or the middle value of the range if it is not. Each data point in the latter distribution is weighted by the product of fraction of total $P(z)$ within the desired redshift range and the fraction of Ly$\alpha$ EW determinations which are below 25\AA, as described in section~\ref{sec:mc}. The $M_\textsc{uv}$ distribution of LAEs covers a broader range of luminosities, and also has a slightly lower median, although the tests reveal no statistical difference. \citet{lemaux_size_2021} use the $5<z<7$ galaxies as part of a larger sample and compare the LAE fraction for the bright ($L \sim 0.67L^*$) and faint ($L \sim 0.1L^*$) counterparts in  and find a higher fraction of LAEs in the faint bin. This may be explained by there being a higher percentage of LAEs among faint galaxies.

We also find there to be no significant difference between the $\beta$ slopes of LAEs and nonemitters, although the median $\beta$ of LAEs is slightly more negative than that of nonemitters. Both simulations \citep{verhamme_3d_2008} and studies at lower redshifts \citep{hayes_redshift_2011, atek_influence_2014} suggest that dust inside of galaxies prevents Ly$\alpha$ photon escape. We therefore initially expect that LAEs may tend to have bluer UV slopes, indicating lower dust content. While we find that the median $\beta$ slope value for LAEs is slightly bluer than that of nonemitters, a KS test indicates no significant difference between the two populations. Figure~\ref{fig:beta} shows the distribution of $\beta$ slopes for the LAE sample and the nonemitter sample, with the same weighting scheme as used in Figure~\ref{fig:muv}. 

We note that the positive $\beta$ slopes for LAEs come from galaxies with few photometric detections with large errors. Those listed for nondetections could be possible low-$z$ interlopers. There is also a possibility of targets in the LAE sample to be low-$z$ interlopers as we have a single line detection, but the chances of this are much lower. We also note that there is sometimes a discrepancy between the $P(z)$ distribution constructed from photometric flux values and the spectroscopic redshift obtained via detection of Ly$\alpha$ emission; \citet{fuller_spectroscopically_2020} show that there is a large ($\sim 35 \%$) outlier rate and that it is not uncommon for the true redshift of a galaxy to lie entirely outside of the $P(z)$ distribution. In targets where this is the case, we note that the $\beta$ slope may not be as reliable due to thin photmetric data and constraints.

For the SED-derived properties of stellar mass, SFR, sSFR, and age, a KS test indicates no statistically significant difference between any of these properties for LAEs vs nonemitters. The distributions can be seen in Figure~\ref{fig:SEDprops}, where the nonLAE properties are weighted using the same prescription as described previously. Between the samples, when employing both a KS and Mann-Whitney test, we do not find any significant difference. The resultant p-values all indicate there is not much evidence that the two samples come from different parent distributions.

\begin{figure*}
\includegraphics[width=18cm]{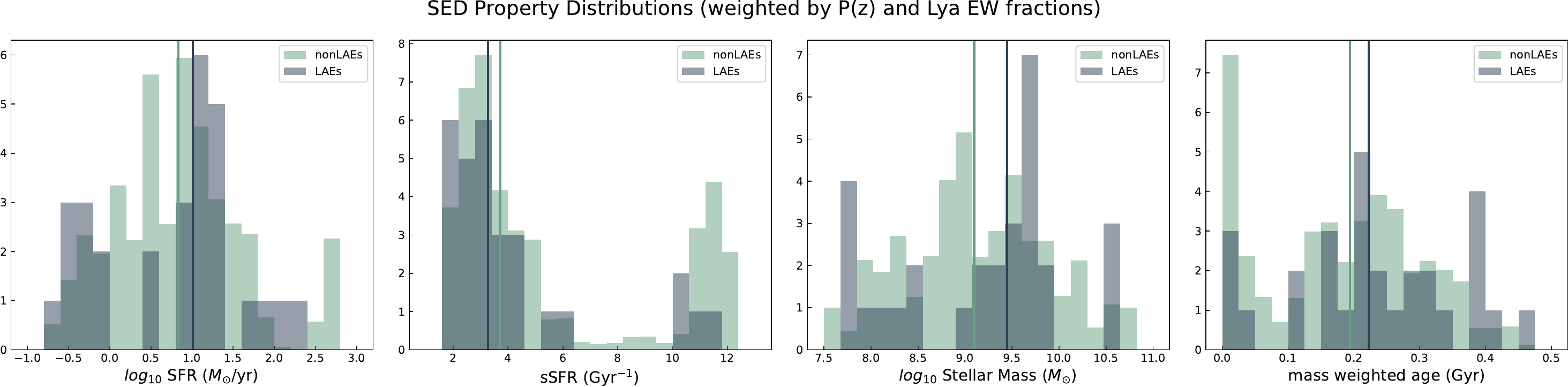}
\caption{Distribution of estimated stellar properties from SED fitting for LAEs vs nonemitters. Solid vertical lines indicate the median value of the distribution with the corresponding color.}
\label{fig:SEDprops}%
\end{figure*}

\section{Spectral Properties of LAEs at $z \sim$ 7}
\label{sec:ciii}
Ly$\alpha$ is typically the brightest UV line in EoR galaxies, making it the first UV emission line that is typically targeted in an attempt to confirm a redshift of such galaxies. However, it is not present in every galaxy and also has a tendency to be scattered by neutral hydrogen, leading to a redshift value that is offset from systemic. When Ly$\alpha$ is detected in galaxies, it is likely to be spectrally offset due to outflows and scattering. In a partially neutral IGM, Ly$\alpha$ is more likely to escape if the galaxy is in an ionized bubble. Due to the redshifting of Ly$\alpha$ from the systemic value, detecting an alternate, non-resonant emission line can help to gain insights into the ISM state of galaxies as well as determine the true redshift \citep[e.g.][]{dijkstra_ly_2014, yang_ly_2017, guaita_vimos_2017, mason_universe_2018, cassata_alpine-alma_2020}. We selected six spectroscopically confirmed LAEs at $z \sim$ 7 to follow up with Keck/MOSFIRE to look for CIII] 1907, 1909 \AA \ emission, whose properties are listed in Table~\ref{table:ciii}. Five of the six are included in all previous analyses, and the sixth, coined Dichromatic Primeval galaxy at $z\sim 7$ (DP7), has a strong Ly$\alpha$ detection described by \citet{pelliccia_relics-dp7_2021}. DP7 is not included in our sample as it was targeted separately in a different survey program, the Reionization Lensing Cluster Survey (RELICS; \citealp{coe_relics_2019}). For all previous analyses, we treat the triply imaged LAE detected in the field of the cluster MACS2129 as one target and do all SED fits on the photometry of the brightest image, Image A. Three full nights were awarded to the project, and we observed for an additional two half nights for a total of 35 hours on sky.

\begin{table*}$z \sim$ 7 LAE Properties
    \begin{center}
    \begin{tabular}{||c c c c c c c c||} 
     \hline
    Galaxy Name & $\alpha_{J2000}$ (deg) & $\delta_{J2000}$ (deg) & $z_{\textrm{Ly$\alpha$}}$ & EW$_{\textrm{Ly$\alpha$}}$ (\AA) &  Ref$^a$ &$t_{\textrm{obs}}$ in H band (min) & EW$_{\textrm{CIII]}}$ (\AA, 3$\sigma$) \\ 
    \hline\hline
    MACS0744-064 & 116.24648 & 39.46042 & 7.148 $\pm$ 0.001 & 58.3 $\pm$ 25 & [1] & 360 & < 43.95\\ 
    MACS1423.16 & 215.928929 & 24.072848 & 7.101 $\pm$ 0.001 & 189 $\pm$ 25 & [2] & 490 & < 28.08 \\
    MACS2129-A & 322.350936 & $-$7.693322 & 6.846 $\pm$ 0.001 & 60 $\pm$ 11 & [3] & 318 & < 20.20 (stacked) \\
    MACS2129-B & 322.353239 & $-$7.697442 & 6.846 $\pm$ 0.001 & 47 $\pm$ 9 & [3] & 318 & ...\\
    MACS2129-C & 322.353943 & $-$7.681646 & 6.846 $\pm$ 0.001 & 170 $\pm$ 77 & [3] & 318 & ...\\
    RXJ1347-018 & 206.89124 & $-$11.75261 & 7.161 $\pm$ 0.001 & 27.2 $\pm$ 7 & [1] & 354 & < 27.38\\
    RXJ1347.47 & 206.900859 & $-$11.754209 & 6.771 $\pm$ 0.001 & 55.4 $\pm$ 10 & [2] & 354 & < 22.91\\
    DP7 & 152.6593385 & $-$12.6556351 & 7.0281 $\pm$ 0.0003 & 237.12 $\pm$ 58 & [4] & 672 & < 47.18\\[1ex] 
     \hline
    \end{tabular}
    \flushleft{$^a$ References for Ly$\alpha$ detections: $[1]$\citet{hoag_constraining_2019}, $[2]$\citet{fuller_spectroscopically_2020}, $[3]$ \citet{huang_detection_2016}, $[4]$\citet{pelliccia_relics-dp7_2021}.} \\
    \caption{LAE Properties of $z \sim 7$ galaxies used in spectroscopic CIII] emission search}
    \label{table:ciii}
    \end{center}
\end{table*}

\begin{figure}
\includegraphics[width=9.2cm]{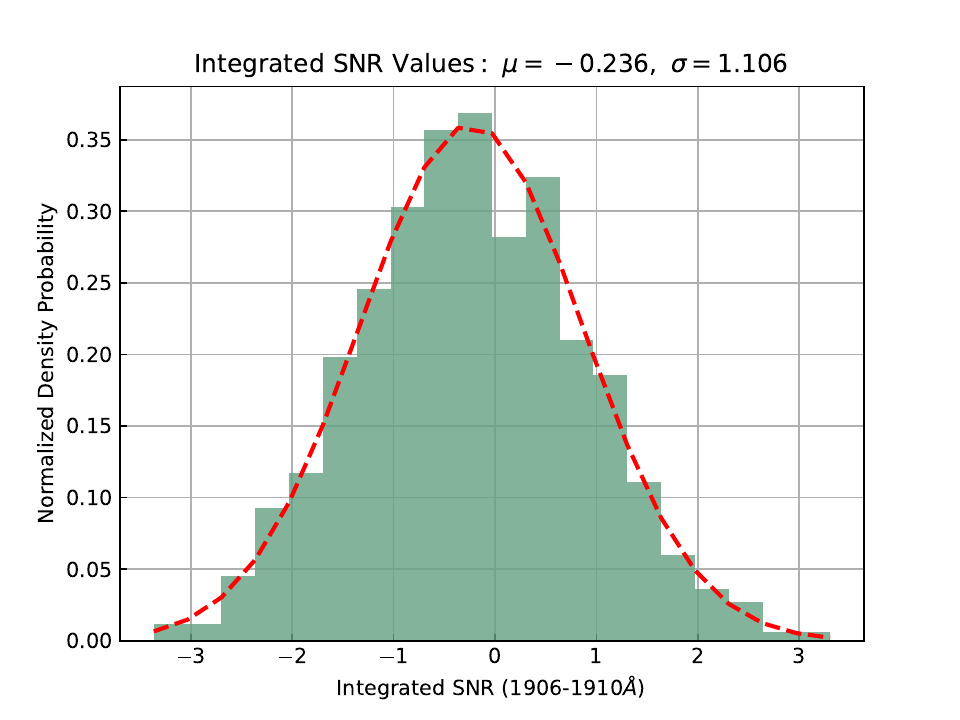}
\caption{Distribution of integrated stacked flux SNR values for the six galaxies used in the CIII] search. These values give no indication of any CIII] emission when stacking the spectra of all the galaxies. The red dashed line shows a Gaussian fitted to the distribution. The slightly negative peak value may indicate modest background oversubtraction during reduction.}
\label{fig:stackedciii}%
\end{figure}

The reduction of the data was done using the pipeline developed by the MOSFIRE Deep Evolution Field (MOSDEF) survey \citep{kriek_mosfire_2015}. This process accounts for any potential instrumental drift as well as any differential atmospheric diffraction by tracking a star in one of the slits, ensuring that minimal signal is lost when combining frames, which is especially pertinent for faint galaxies. After recovering no obvious emission lines in a by-eye search from the initial reductions, we stress test the frames to guarantee that no signal was missed. These tests include varying the width of the boxcar used in the 1D extraction, removing frames where there was cirrus cloud coverage, and performing an automated integrated signal to noise ratio search in addition to probing the reduced spectra by eye. After testing each reduced galaxy meticulously, we confirm that there is no CIII] emission detected from any of the targeted galaxies. While there is always a possibility of emission lines being obscured by atmospheric skylines when using ground-based near-infrared spectra, the chances of both lines in the CIII] doublet being obscured is low. However, we cannot rule out that this may have happened.

We also perform a test for faint CIII] emission by stacking the spectra of all six LAEs to see if there is any significant signal to noise recovered. In this process, we use the Ly$\alpha \ \Delta v$ distribution from \citet{cassata_alpine-alma_2020} as CIII] should trace the systemic velocity of the galaxy \citep[e.g.][]{talia_gmass_2012, talia_agn-enhanced_2017}. We sample an offset from this distribution for each of the six LAEs, shift the spectrum by that amount, and perform an inverse variance weighted stacking at that location. Over all iterations, we recover no appreciable integrated signal-to-noise at the expected location of CIII]. Figure~\ref{fig:stackedciii} shows the distribution of the stacked integrated signal-to-noise ratio (SNR) across the spectral window where CIII] would be expected for 1000 iterations. 

\begin{figure}
\includegraphics[width=9.2cm]{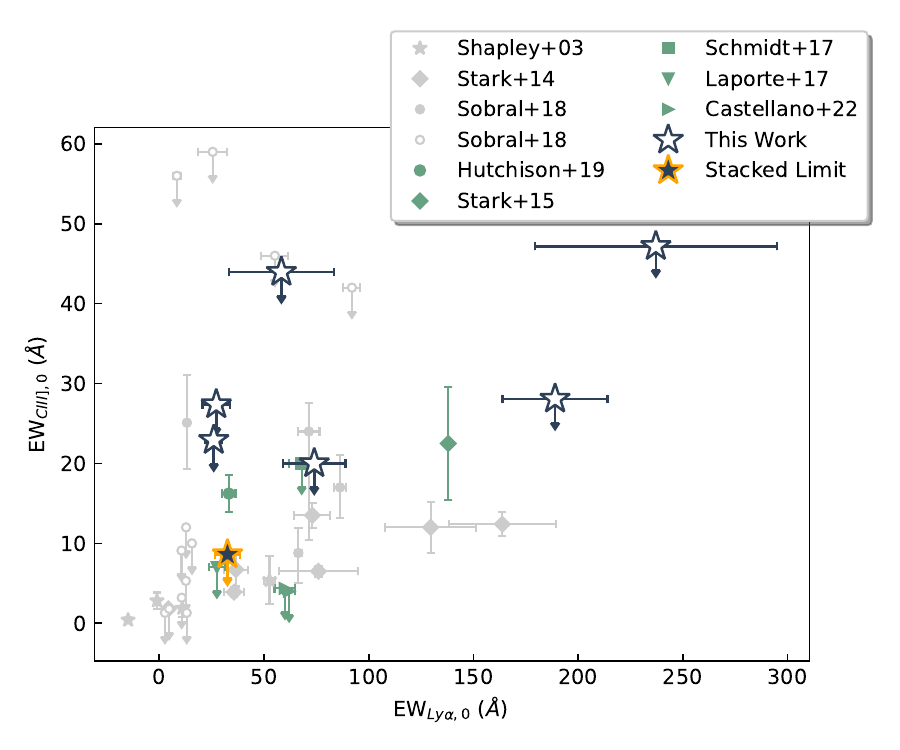}
\caption{Ly$\alpha$ vs CIII] EW value or upper limits for LAEs }
\label{fig:ciii}%
\end{figure}

Even with no detected CIII], we can use the upper limits on the relative strength of the individual lines as well as the stacked limit to gain insights into these LAEs and rule out certain characteristics. For each galaxy targeted, we compute the 1$\sigma$ upper limit on the rest-frame EW of CIII] using the same process as described in section~\ref{sec:galprops}. The 3$\sigma$ upper limits on the EW of CIII] are noted in Table~\ref{table:ciii} and plotted in Figure~\ref{fig:ciii}. We calculate a stacked EW limit on CIII] of 8.6\AA \ at the 3$\sigma$ level for all of the galaxies together. 

Typical CIII] EW values from the $z \sim 2-3$ universe are less than 15\AA \ with a median value around 7\AA \ \citep{stark_ultraviolet_2014, llerena_vandels_2022}. Sources at higher redshifts however have detected CIII] emission at higher EW values. \citet{stark_spectroscopic_2015}, \citet{hutchison_near-infrared_2019}, and \citet{topping_detection_2021} classify CIII] emission lines at $z = 6.027, z = 7.51,$ and $ z = 7.945$, and  with EW values for the combined doublet of 22.5$\pm$ 7.1\AA, 16.23$\pm$ 2.32\AA, and 20.3$\pm$ 6.5\AA, respectively. Recent NIRSpec observations by \citet{tang_jwstnirspec_2023} find CIII] emission in three galaxies at $7.8<z<8.7$ with EW values ranging from 10.9-16\AA.  At high redshifts, there is evidence of stronger CIII] emission than in lower-$z$ galaxies. In addition, observations from the lower redshift universe have shown a correlation between Ly$\alpha$ EW and CIII] EW \citep{llerena_vandels_2022}. Considering these, we may expect some high EW CIII] emitters in our $z \sim 7$ subsample, but we do not and constrain the 3$\sigma$ stacked EW limit to 8.6\AA, a lower value than any of the high-$z$  detected  CIII] lines cited above. 

Photoionization modelling of confirmed strong CIII] emitters reveals certain properties that allow for the increased production of doubly ionized carbon \citep[e.g.][]{hutchison_near-infrared_2019}. Observations of strong CIII] in lower redshift galaxies have found that such measurements are found in galaxies with low metallicity, high ionization parameters, and hard ionizing spectra \citep{erb_physical_2010, steidel_reconciling_2016, du_searching_2020, tang_rest-frame_2021}. \citet{hutchison_near-infrared_2019} use their detection as well as limits on other nebular emission lines to determine that the galaxy in study has subsolar metallicity, a high ionization parameter, and a young stellar population. Comparing to models presented in \citet{nakajima_mean_2018, nakajima_vimos_2018}, we cannot rule out AGN activity, but galaxies with low CIII] EW limits, like some of those in our subsample, are more likely to be star-forming galaxies than have AGN activity, and also less likely to have extremely high ionization parameters. This is perhaps surprising given that some of the targeted sample has very strong observed Ly$\alpha$ and the similarity of the strength of the Ly$\alpha$ emission relative to other EoR galaxies that have strong observed CIII] emission.

\section{Discussion}
\label{sec:discussion}
Perhaps the most interesting results from this study are those presented in Sections~\ref{sec:EWvsphys} and~\ref{sec:LAEvsnon}: the lack of any significant correlations or differences where they may be expected from previous studies. We find no significant differences between any of the distributions of properties of LAEs and LBG candidates without Ly$\alpha$ emission, as well as no significant correlation between Ly$\alpha$ EW and physical properties for those with spectroscopic detections. Out of the numerous studies which look at the physical differences between LAEs and nonemitters in order to pinpoint what about the emitters allows for the escape of moderate to strong Ly$\alpha$ emission, many studies generally find a consensus that LAEs are bluer, less massive, less star-forming, and fainter. There is scatter within these results, however. Below we discuss both physical and statistical reasons which may explain our results.

Our conclusions may be different than those of other surveys likely due to two properties of our sample which set it apart from others: high redshifts and faint UV luminosities. Most other studies which perform similar analyses at faint luminosities are at redshifts of $z < 6$. Galaxies in the early universe are more difficult to study and characterize and may be fundamentally different than those at similar stages of evolution at lower redshifts, largely due to their light being at least partially obscured by the opacity of the IGM. 

The environment in which these galaxies are formed and existing is a key facet to consider. The IGM is at least partially neutral at $z \gtrsim 6$, with lingering patches of neutral hydrogen possibly present at lower redshifts \citep[e.g.][]{stark_keck_2010, mason_universe_2018, bolan_inferring_2022}, potentially affecting each galaxy in this study. As neutral hydrogen absorbs the resonant Ly$\alpha$ line, the EW values we obtain may be artificially lower than what has intrinsically escaped the galaxy. Due to the patchiness of reionization \citep{furlanetto_effects_2006, treu_changing_2013, sobacchi_inhomogeneous_2014}, this effect may not be homogeneous across all galaxies in differing fields of view. A similar consideration is the evidence of overdensities of galaxies at the same redshift within a small physical region, which have been found at $z > 6$ \citep{castellano_spectroscopic_2018, larson_searching_2022, jung_new_2022, morishita_early_2023, hashimoto_rioja_2023, cooper_web_2023}. Perhaps galaxies which reside within these overdense regions have more of a chance of their Ly$\alpha$ photons escaping into the IGM due to the expanse of the ionized bubbles in which they reside. We look at the redshift distribution of LAEs within each cluster to look for evidence of overdensities in our sample. Upon further inspection of the colors and lens models, We find that there may a lensed pair, with detections in the RXJ1347 field placing two galaxies at redshifts of 5.194 and 5.195. We check all of the correlation tests including just the brightest target in this possible double image, and the results remain unchanged. We do not see any strong evidence of possible protoclusters in our sample, but their existence cannot be ruled out. As our study probes 10 lines of sight, this effect is mitigated, but could still have some impacts. The side effects of an evolving IGM and patchy overdensities may cause our EW values to be affected sporadically, leading to the lack of strong correlations between Ly$\alpha$ EW and physical properties. In this case the relationships between physical quantities and Ly$\alpha$ EW could be changed or even completely erased by modulation from the environment. However, we note that \citet{jones_jades_2023} studies galaxy samples that extend into the EoR up to $z \sim 8$ and do find anticorrelation with Ly$\alpha$ EW and UV luminosity and dust extinction, respectively, but these samples do not extend quite as far into the faint luminosity regime. Additionally, \citet{jones_jades_2023} find that their anticorrelation may not be an intrinsic property of the sample, but possibly due to flux sensitivity limits.

We test if modulation by an opaque IGM is a significant factor by looking at how the Ly$\alpha$ EW trends for emitters change when only including galaxies at $z \leq 6$, when the Universe was likely mostly ionized \citep[e.g.][]{fan_observational_2006}. We find that the significance of the expected anticorrelation between EW and stellar mass and SFR increases, to the $\sim 1.5\sigma$ level. While this is still not strong enough to confidently claim a correlation, it is perhaps indicative that the increasingly neutral IGM at redshifts above $z \sim 6$ may play a role in our results.

In addition to the environment in which the galaxies are forming, this sample is unique in that it is comprised of LBG candidates which are characteristically faint ($< L^*$), observable due to magnification from massive lensing clusters. The $M_\textsc{uv}$ distribution of our sample is much fainter than that of any other surveys to which we compare; the typical value of the intrinsic absolute UV magnitude of the galaxies in our sample is $\sim -19$, reaching down to $M_\textsc{uv} \sim -14$, whereas other comparison surveys generally probe galaxies closer to $L^*$, at absolute magnitudes of $-22 \lesssim M_\textsc{uv} \lesssim -18$. We are exploring unchartered phase spaces in this study, as there have been no surveys at these redshifts with galaxies as faint as those in our sample.\citet{saxena_jades_2023} also find a high-EW ($\sim 300$\AA) LAE at $z=7.3$), at $M_\textsc{uv} \sim -17$, similar to galaxies in our sample, but also do not probe galaxies fainter than this. As noted in \citet{mccarron_stellar_2022}, low-mass, low-EW systems are notoriously hard to study, so often samples are comprised of more rare bright LAEs, which may not be representative of the typical galaxy population. While our survey is certainly not insusceptible to Malmquist bias, gravitational lensing does allow for some mitigation of its effects. We detect these types of galaxies that are missed in field surveys (as due to lensing, we can achieve the same EW sensitivity for a faint, lensed object as for a bright one without lensing), but whose physical properties veer away from the tight correlations that are presented in other works, as can be seen in the bottom center panel of Figure~\ref{fig:LAEprops}. 

%expand this and show LAE fraction plots maybe? talk about completeness?
Our lack of correlations among Ly$\alpha$ line strength of LAEs and their physical properties may be not only due to the faint galaxies we detect with low Ly$\alpha$ EW, but also the large scatter we see at the high luminosity end of our sample. We see both our highest and lowest Ly$\alpha$ EW at bright luminosities. Our sample is characteritstically faint, and we do not have a large number of galaxies brighter than $M_\textsc{uv} \sim -21$. Many other samples cover brighter UV luminosities; perhaps if brighter targets were probed, we would recover the low EW values seen in other surveys and see correlations or more difference between the LAE and nonemitter populations. These results may suggest that the fainter population of galaxies are inherently different from their bright counterparts as they do not exhibit the same significant trends. 

%throw this guy in somewhere here
% Our disparate results in comparison to other surveys may indicate that there is no Ly$\alpha$ EW dependence on $M_\textsc{uv}$ toward fainter luminosities. However, our sample also shows significant scatter in EW values toward the bright end of the luminosity distribution. The addition of data points in this section of phase space can weaken the correlations seen in other surveys, which we discuss further in Section~\ref{sec:discussion}. In addition, the relationship between Ly$\alpha$ EW and UV luminosity may not be straightforward due to dependence on the escape fraction of Ly$\alpha$ ($f_{\textrm{esc, Ly$\alpha$}}$) or redshift-dependent ionizing efficiency \citep{matthee_spectroscopic_2017, sobral_slicing_2018}. 

Lastly, we also consider our sample size. While we have spectroscopic data on 247 galaxies and LBG candidates, we have 38 which are confirmed to be at $5 < z < 8.2$ from Ly$\alpha$ emission, and the rest have photometric redshifts with varying degrees of certainty to be within the range where we could detect Ly$\alpha$ emission. When we weight each nomemitter using the scheme defined in section~\ref{sec:mc}, the size of the effective nonemitter sample is comparable to that of the emitters. Perhaps the intrinsic scatter shrouds any correlations or population differences with a relatively small sample size. We note that the uncertainty in redshift for the nonemitters may affect these results; however, those LBG candidates which are most likely to be at the redshifts we are studying are the ones which make the most contribution to the comparison between LAEs and nonLAEs. 

We perform tests to see if significant correlations that are evident in a large sample would be detectable with our sample size. First, we take the data from the VIMOS Ultra Deep Survey (VUDS; \citealp{le_fevre_vimos_2015, tasca_vimos_2017, lemaux_vimos_2022}), a spectroscopic survey of galaxies with $0.3 < L_{\textrm{UV}}/L^* < 3$ over the redshift range $2 < z < 6$, which has spectroscopic observations that are integrated for long enough to reach the continuum, such that Ly$\alpha$ in emission is not requirement for a redshift. These which show weak ($\rho = -0.18$) anticorrelation between Ly$\alpha$ EW in emission and SFR at $> 5\sigma$ significance at $2 < z < 6$. From this data, we subsample a random draw of 38 galaxies and compute the Spearman correlation statistic and p-value between Ly$\alpha$ EW and SFR. Over all iterations of this subsampling, only $\sim 3 \%$ show a $>3\sigma$ significance. We note that we are not directly comparing the results of this survey with ours as they are at different redshift ranges, but simply using the sample to see if weak correlations found in large samples would be evident in one of our size. These results show that we may not recover a similarly weak trend with our sample size. 

We apply the same method to the data presented by JADES \citep{jones_jades_2023}, whose original sample combined with data from literature in the same redshift space show strong correlation ($\rho = 0.65$) between Ly$\alpha$ EW and $M_\textsc{uv}$ at $> 5\sigma$ significance at a similar redshift range to ours, $3 < z < 8$. When we perform the subsampling, all iterations result in strong correlations. All realizations return significances of $>2\sigma$, and the vast majority show a $>3\sigma$ significance. These exercises show that if a strong correlation exists in the underlying population of galaxies, a sample of our size would likely show that correlation. However, as we do not find any evidence of correlation between Ly$\alpha$ EW and any of the physical properties studied, we conclude that, if there does exist a trend among the underlying population of galaxies we probe, it is weak at best. 

Our lack of expected correlations or differences between LAEs and nonemitters among sub-$L^*$ galaxies is likely due to some combination of environmental effects, specifically inhomogeneous IGM opacity, and a fainter sample than is typically studied. It is unlikely to be solely due to having a low sample size $-$ if a strong correlation exists among an underlying population, it would still likely show at high significance in a small subsample. Our inclusion of low Ly$\alpha$ EW, low luminosity galaxies as well as bright, large EW galaxies may reduce correlation strength in the properties we study, and also contribute to the similar distributions among LAEs and nonLAEs. The correlations between Ly$\alpha$ EW and physical properties such as stellar mass, SFR, UV luminosity, and $\beta$ slope, appear to be weaker among faint $z > 5$ galaxies than some other samples suggest. 

%other studies have shown large scatter in the properties of their galaxies and find that LAEs and LBGs are not monolithic \citep{pentericci_physical_2009, nilsson_nature_2011, jung_ceers_2023}.

\section{Conclusions}
\label{sec:conclusions} 
We have analyzed the stellar and UV properties of a sample of 247 faint, gravitationally lensed LAEs and LBG candidates between $5 < z < 8.2$ with both deep photometric and spectroscopic data. We investigate how Ly$\alpha$ EW correlates with UV and stellar properties for galaxies with detected emission, and we also compare the distributions of these properties for LAEs and the nonLAE sample. 
\begin{itemize}
\item We do not find any significant correlations between Ly$\alpha$ EW and any of the stellar or UV properties which we analyze in this work. 
\item We find no significant difference between the stellar and UV properties of LAEs and nonemitters.
\item Our lack of correlations indicate that if trends exist in this population of faint EoR galaxies, they are weak. We detect low-EW, faint galaxies which other surveys do not, and also have high scatter in properties at the bright end of our sample. Our results could also be modulated by suppressed Ly$\alpha$ emission due to a partially neutral IGM at these redshifts. Any weak trends could be rendered less significant by the uncorrelated opacity of the surrounding IGM.
\item We do not find CIII] emission in a spectroscopic search of six confirmed LAEs at $z \sim 7$ and calculate their EW upper limits, some of which are constraining enough to rule out extreme nebular emission properties and disfavor AGN as being the dominant source powering emission.

Continued observations of high-$z$ galaxies with $JWST$ will allow for large sample sizes of EoR galaxies with confirmed redshifts and the necessary data to confidently estimate physical properties via SED fitting, as well as accurate $\beta$ slopes from spectra. Even in the era of $JWST$, large samples of ground-based observations are important for detecting low EW Ly$\alpha$ lines at these redshifts, as it is difficult even with the NIRSpec prism. Taking advantage of these and future observations in the coming years will narrow down the properties which drive Ly$\alpha$ and ionizing photon production and escape in EoR galaxies, allowing for the characterization of the sources which caused cosmic reionization. 

\end{itemize}

\section*{Acknowledgements}
We acknowledge support from the program HST-GO-16667, provided through a grant from the STScI under NASA contract NAS5-26555. MB acknowledges support from the ERC Grant FIRSTLIGHT and Slovenian national research agency ARRS through grants N1-0238 and P1-0188.  Based on spectrographic data obtained at the W.M.Keck Observatory, which is operated as a scientific partnership among the California Institute of Technology, the University of California, and the National Aeronautics and Space Administration. The Observatory was made possible by the generous financial support of the W.M. Keck Foundation. The authors wish to recognize and acknowledge the very significant cultural role and reverence that the summit of Maunakea has always had within the indigenous Hawaiian community.  We are most fortunate to have the opportunity to conduct observations from this mountain. Also based on observations made with the NASA/ESA Hubble Space Telescope, obtained at the Space Telescope Science Institute, which is operated by the Association of Universities for Research in Astronomy, Inc., under NASA contract NAS5-26555. And based on observations made with the Spitzer Space Telescope, which is operated by the Jet Propulsion Laboratory, California Institute of Technology under a contract with NASA. Support for this work was also provided by NASA/HST grant HST-GO-14096, and through an award issued by JPL/Caltech. GCJ acknowledges funding from the "FirstGalaxies" Advanced Grant from the European Research Council (ERC) under the European Union’s Horizon 2020 research and innovation programme (Grant agreement No. 789056). PB thanks Spencer Fuller, Austin Hoag, Kuang-Han Huang, and Charlotte Mason for their immense contributions to this work, having designed and executed many of the observations, data reduction, and preliminary analyses. We thank the referee, Aayush Saxena, for their comments and insights which greatly improved this paper and for their kindness at the conclusion of this process. 
\\

\noindent This study was performed and written, in between adventures, by Patricia (Patty) Marie Bolan, who passed away shortly before its final submission. This paper is dedicated to Patty's brilliant life and light. Patty was a gifted astrophysicist, passionate science communicator, dear friend, and beloved daughter and sister. The stars shine less brightly for her absence. 

%%%%%%%%%%%%%%%%%%%%%%%%%%%%%%%%%%%%%%%%%%%%%%%%%%
\section*{Data Availability}
To the best of our ability, the data underlying this article will be shared upon reasonable request to the corresponding author.
%A table of all the LAEs and galaxy candidates used in this analysis, along with all the physical properties examined, can be found on CDS. 



\bibliographystyle{mnras}
\bibliography{Lya_bib_2} % if your bibtex file is called example.bib

%\begin{thebibliography}{99}
%\end{thebibliography}

%%%%%%%%%%%%%%%%%%%%%%%%%%%%%%%%%%%%%%%%%%%%%%%%%%

%%%%%%%%%%%%%%%%% APPENDICES %%%%%%%%%%%%%%%%%%%%%

%\appendix

%%%%%%%%%%%%%%%%%%%%%%%%%%%%%%%%%%%%%%%%%%%%%%%%%%

% Don't change these lines
\bsp	% typesetting comment
\label{lastpage}
\end{document}